\documentclass[aps,twocolumn,floats,floatfix,prl,psfig]{revtex4}
\usepackage{graphicx,amssymb,amsmath,times}
\newcommand{\be}{\begin{equation}}
\newcommand{\ee}{\end{equation}}
\newcommand{\beq}{\begin{equation}}
\newcommand{\beqa}{\begin{eqnarray}}
\newcommand{\eeq}{\end{equation}}
\newcommand{\eeqa}{\end{eqnarray}}

\newcommand{\dg}{$^\circ$~}

\begin{document}
\title{Massive galaxy clusters and the origin of Ultra High Energy Cosmic Rays }
  
\author{Elena Pierpaoli}
\email{pierpa@caltech.edu}
\affiliation{Theoretical Astrophysics, MC 130-33, California Institute of Technology, Pasadena,
CA 91125
}

\author{Glennys Farrar}
\email{gf25@nyu.edu}
\affiliation{Center for Cosmology and Particle Physics, 
Department of Physics, New York University, 
New York, NY 10003, USA
}
\begin{abstract}
We investigate whether ultra--high energy cosmic rays (UHECR) may be preferentially produced in 
massive galaxy clusters, by looking for correlations between UHECR directions and those of x-ray clusters. 
We find an excess-above-random of high energy cosmic rays which correlate with massive galaxy cluster positions. 
For cosmic rays with energies above 50 EeV the observed correlation is the strongest or angles of
1.2-1.6 degrees where it has a chance probability of about  0.1  \%.
 Including lower energy cosmic rays in the sample causes the angle where the most significant correlation is found to increase, as would be expected by virtue of instrumental and magnetic smearing increasing at lower energy.   
 These results suggest that some UHECR are produced in galaxy clusters, or in objects that
 preferentially populate galaxy clusters.
 
\end{abstract}
\maketitle

\underline{\em 1)  Introduction}
The origin of the most energetic particles in nature, ultrahigh energy cosmic rays (UHECR) with EeV energies and above, is one of the outstanding mysteries in astrophysics.  Lacking a convincing theory or identifiable source, their existence has even been attributed  to  very diverse physical phenomena like the decay of 
super--heavy  particles \cite{CMR04}, or particle acceleration in  powerful extragalactic objects with particular attention dedicated to BLLac objects \cite{VDG03,GTT02,HR:BLLacs} . Such speculations have raised a lot of discussions \cite{ EFS03,EFS04,TT04},  
leaving the question 
of the origin of UHECR still open.

Our purpose in this paper is to investigate another potential source of UHECRs which has been highlighted recently, namely massive clusters of galaxies.  The motivation for our study is the observation\cite{fbh05} that a cluster of 4 events in the combined AGASA-HiRes high energy dataset\cite{HRGF}, which becomes 5 events in the entire published HiRes-AGASA dataset\cite{gfclus}, is well-consistent with coming from a merging pair of galaxy clusters identified at a distance of 140/h Mpc in the SDSS DR3, while no other particularly striking objects, such as powerful radio galaxies or BL Lacs, are found within the estimated error radius for the source direction\cite{fbh05}.  The probability for finding as strong a clustering signal by chance is of order half a percent\cite{HRGF, gfclus}, but verification with new data is needed for it to be considered a definitive cluster signal.   A merging pair of clusters would be expected to have very large scale, strong magnetic shocks which could be responsible for accelerating UHECR even if there is no AGN or GRB associated with the galaxy clusters\citep{fbh05}. 

 Here we investigate the possibility that other systems which can have comparably large shocks -- very massive clusters of galaxies, where shocks are generated by infalling matter -- may also be linked to UHECR generation.  To that end, we look for correlations between possible UHECR source directions and massive galaxy clusters, using the method of \cite{TT:BLLacs}.  The correlation we find justifies a more refined analysis using the maximum likelihood method employed in\cite{HR:BLLacs}, however that requires unpublished information about the HiRes stereo exposure.

\underline{\em 2)  The x-ray cluster data} \label{xraycat}
In order to study the possible correlation of UHECR directions with clusters, we need a sample of galaxy clusters homogeneusly selected on the vastest possible area covered by the UHECR experiments.  To this aim, we use the extended X-ray selected galaxy catalog NORAS  \cite{Bo00}. The NORAS catalog covers the northern celestial hemisphere $\delta \ge 0 ^{\circ}$  with a restriction on the galactic latitude:   $|b| \ge 20^{\circ}$. We further restrict to the area of the sky which overlaps with the region of good spatial coverage of the AGASA and HiRes UHECR detectors: ${\delta}  \le 80^{\circ}$.

\begin{figure}[ht] 
   \begin{center}
  \includegraphics[width=2.8in]{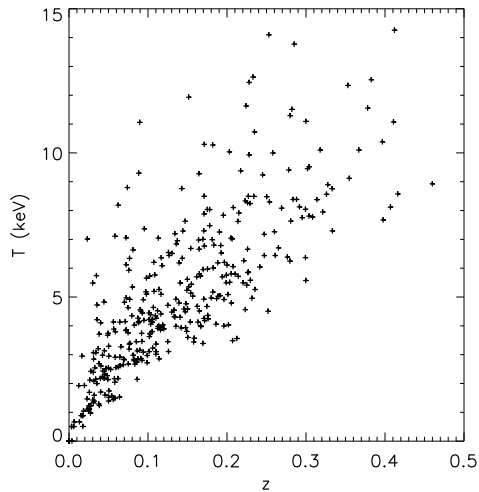} 
   \end{center}
   \caption{ The distribution of the NORAS x-ray clusters in the redshift -- temperature plane. }
   \label{fig:clucat}
\end{figure}

The main NORAS catalog contains 378 objects  (375 in the area considered here)  selected for having extended X--ray emission in ROSAT All-Sky Survey.  Given the restrictive selection criteria, the catalog is likely to be highly pure but is  thought to be only about 50\% complete for fluxes above $ 3 \times 10^{-12} {\rm erg s}^{-1} {\rm cm}^{-2}$. 
The redshift distribution of the clusters in the main catalog is reported in figure \ref{fig:clucat} where we have evaluated the temperature according to  Ikebe et al \cite{Ik02}.
An additional 18 x-ray clusters are listed separately by NORAS because they may be aligned with AGN, for which no redshift estimate is available.  We take as our primary sample the extended catalog of 393 clusters: 375 clusters from the main catalog plus 18 from the extention, in the overlap region with the UHECR exposure. 

\underline{\em 3)  The cosmic ray data} \label{uhecrs}
We will discuss here (separately) the correlations of x-ray clusters with UHECR events reported by two different experiments, AGASA and HiRes, excluding those arriving from directions with galactic latitude $|b| <20^\circ$ and RA $\delta>80^\circ$ where the UHECR exposure is non-uniform.  For AGASA\cite{AGASAupdate} we consider two samples, the full published AGASA dataset of events above 40 EeV (AG40), and just events above 50 EeV (AG50). The total number of UHECR directions we have in these samples, after the fiducial cuts in angle, is 36 and 21, respectively.  The HiRes stereo experiment has published coordinates of 271 events above 10 EeV\cite{HRclus04}, of which 143 events (the "HR" dataset) are in our fiducial region.

For each AGASA direction we assign a ``GZK maximum distance" obtained using results of Elbert and Sommers as follows.  Fig. 2 of \cite{ES95} gives for several values of observed CR energy, the locus of source distance versus the probability the observed CR could have been produced beyond that distance, calculated assuming a source spectrum $dn/dE \sim E^{-2.5}$ and including the effects of photopion and pair production.  For instance, a CR observed with 80 EeV has a probability of 0.8\% of having a source further than 200 Mpc.  A quadratic interpolating function to this plot is used to solve for the distance $d_{99}(E)$ such that an event of a energy $E$ has no more than a 1\% probability of having originated further than the quoted distance.  The combined systematic and statistical error of AGASA energies is 30\% \cite{AGASAenergy} so we define the ``GZK Maximum Distance" of an event to be $d_{99}(0.7 E)$ unless it is larger than 1 Gpc.  In that case, we impose no constraint because our crude method of estimating $d_{99}$ is unreliable for very large distances.  Altogether, 7 UHECR directions have a GZK distance constraint.  HiRes published only coordinates and not energies so we cannot impose a GZK maximum distance constraint for the HR dataset. 

\begin{figure}[ht] 
   \begin{center}
  \includegraphics[width=2.8in]{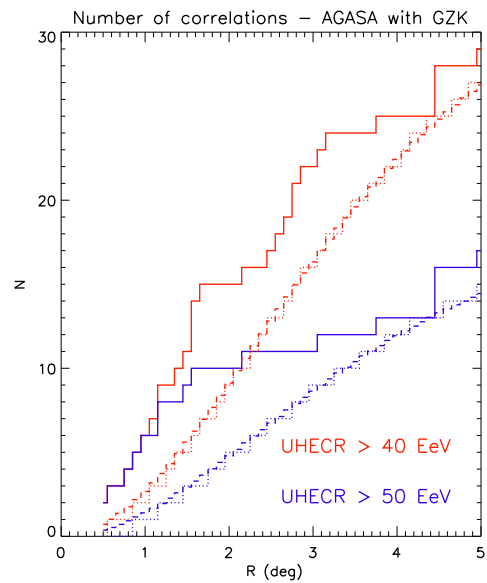} 
   \end{center}
   \caption{ Observed number of correlations as a function of radius for AG50 and AG40 (solid blue and red lines).  Mean and most probable number of correlations in corresponding randomly distributed CR datasets (dotted and dashed lines). }
   \label{fig:rvc}
\end{figure}

\underline{\em 4)  Method and Results} \label{meth}
For each set of  UHECR directions and maximum distances, we find the nearest of 
the 393 NORAS x-ray clusters to the CR direction and tabulate the correlations in radial bins of 0.1 degrees.  Figure \ref{fig:rvc}  shows such results 
for the two AGASA CR sets. Imposing the GZK distance constraint does not change the number of correlations for angles below 1.4 degrees. Note that inclusion of CR between 40 and 50 EeV only increases the number of correlations at radii greater than one degree.  The parameters of the seven correlated x-ray clusters and associated CR whose angular separations are $\le 1.2^\circ$ are given in Table \ref{tab:cluc}.

\begin{table}[h]
\centering

\begin{tabular}{cccccc}

   E (EeV)  &   z  &       T (keV) & ra ($^\circ$) & $\delta$ ($^{\circ}$) &  R($^\circ$)\\
   144.00  & 0.0326  & 2.5  & 241.24 &23.92  & 0.97 \\
   91.00 &  0. & 0.  & 286.43 & 78.08 &   0.87 \\
   77.60  & 0.2065 & 7.0 & 168.60 & 58.39  & 0.76 \\
   72.10  & 0.0763 & 3.4 & 167.33 &41.56  & 0.23 \\
   69.30  & 0.1138  &5.3 &  259.54 &56.66  & 0.46 \\
   61.90   &  0.2080 & 8.1 &   69.76 & 5.34   &0.56 \\
   56.80  & 0.2836  & 8.4 &  198.78 &51.82 &1.17 \\
   55.00 &  0.0527 &  2.5 &173.08 & 55.99  &1.19 \\
49.70 & 0.1813 & 7.0 & 215.42 & 37.30 & 1.01\\
\end{tabular}
\caption{AGASA events nearer to an x-ray cluster than 1.2$^\circ$.
The first column is the energy of the UHECR, followed by correlating cluster's redshift, temperature, coordinates.  The last column is the anglular separation between UHECR and cluster direction.}
\label{tab:cluc}
\end{table}

\begin{figure}[t] 
   \begin{center}
   \includegraphics[width=2.8in]{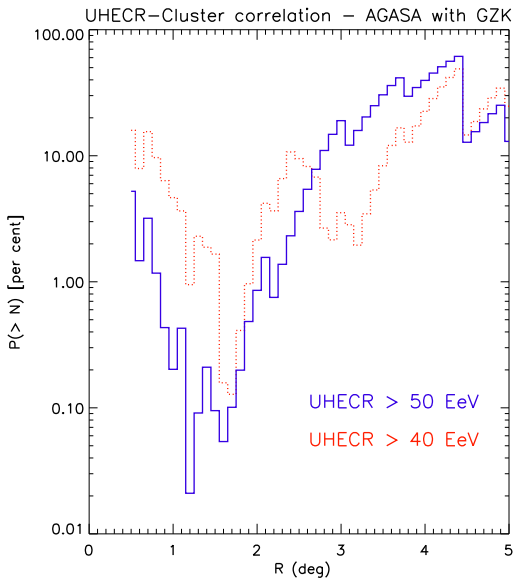}
   \end{center}
   \caption{The probability of finding a number of correlations between AGASA events and x-ray clusters equal or greater than that found in the real data, as a function of the searching radius. The red dotted (blue solid) line corresponds to the  AG40 (AG50) data sets.}
   \label{fig:extGZK}
\end{figure}

In order to assess the significance of our results, we make Monte Carlo realizations considering a hypothetical set of UHECR with the same set of GZK distances as in the real data but randomly chosen directions.  We take the exposure to be uniform which is a good approximation for  $\delta \le 80^\circ$ and in any case does not induce any {\it a priori} bias in the estimate of the chance likelihood.  We then search for correlated x-ray clusters in the same way as for the real data.  Figure ~\ref{fig:rvc} shows the number of correlations observed, and also the number expected in a random sample, as a function of radius. The number of correlations found in the real data substantially exceeds the number in the random sample at all but the largest radii.  
Figure ~\ref{fig:extGZK} gives the probability of finding in a random dataset, an equal or higher number of matching clusters as observed in the real data, as a function of angular separation chosen.  

The most significant correlation for the AG50 dataset occurs in the angular range of about 1.2-1.6 degrees, for which the likelihood of finding the observed number of correlations by chance is of order 0.1\%, with the minimum value being 0.02\% at 1.2\dg.  For the AG40 dataset, the most significant correlation is at about 
1.7\dg  for which the chance probability is 0.13\%.  If the GZK distance constraint is not imposed, the chance probabilities increase by about one order of magnitude.
  
We checked whether any of the UHECR-x-ray cluster correlations below 2\dg  involve UHECR events in the five AGASA doublets or the triplet.  Following the nomenclature of \cite{AGASAupdate}, only  C1, C2 and C6 have more than one event   in our angular region.  C1 events are not correlated with a NORAS x-ray cluster (although one is  with a non-NORAS bright cluster), the C2 events are all correlated but with different clusters, 
and the C6 events are both correlated with the same cluster but just one at an angle below 2\dg.
Only retaining the most energetic event from each CR doublet or triplet doesn't appreciably change the correlation statistics above.

\begin{figure}[h] 
   \begin{center}
   \includegraphics[width=2.8in]{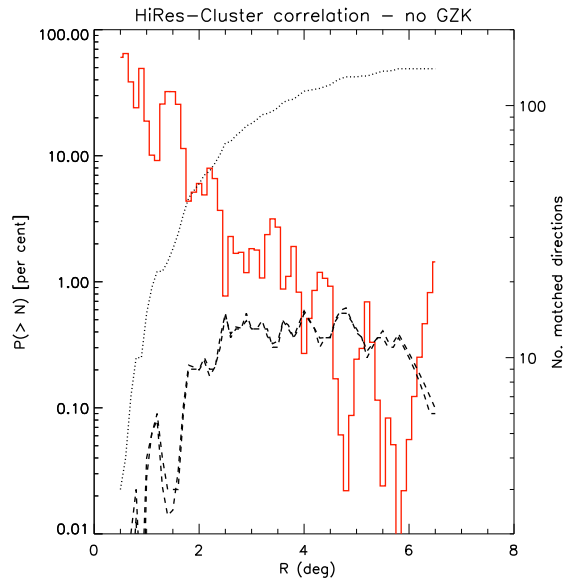}
   \end{center}
   \caption{Correlations between x-ray clusters and HiRes UHECRs above 10 EeV.  Black (scale on right vertical axis): dashed lines are the excess number of correlations compared to the mean and peak of the random distribution and dotted line is the observed number of correlations. Red (scale on left vertical axis): the probability in percent of finding a number of matched clusters equal or greater than the one found in the real data as a function of the searching radius.  }
   \label{fig:HR}
\end{figure}

The HiRes data is complementary to AG40 and AG50 because its average energy is lower and its angular resolution is better than AGASA so magnetic smearing, which $\sim E_{CR}^{-1}$ for a given source, should play a relatively more important role in setting the radius of correlations.   Results of the correlation analysis between the HR events and the extended NORAS catalog are given in Fig. \ref{fig:HR}.  There is an excess above the expected number of random correlations of about 10--15 cases for the 143 HiRes UHECRs, rather independently of angle for angles above about 2$^\circ$.  The probability of finding so large an excess number of correlations by chance is 1\% at about  2\dg where the excess number plateaus.  The continued decrease of the chance probability as the search radius increases is related to the clustered nature of the x-ray clusters themselves.

\underline{\em 5)  Discussion and Conclusions}
The frequencies of finding the observed or greater number of correlations in a random dataset, shown in Figs. 3 and 4, cannot be treated as {\it a priori} probabilities, because the most sensitive angle is not {\it a priori} known, since the extent of magnetic smearing is not known.  Also, the result depends on how the data is binned.  If one could use the assumption of no magnetic smearing, these problems could be eliminated by doing an unbinned maximum likelihood analysis using the actual resolution of the events, as in refs. \cite{HRGF, HR:BLLacs}.  The assumption of no magnetic smearing is not justifiable, but we can sidestep the sensitivity to binning and to the fact that many radii are being inspected by calculating how often -- in a large number of random datasets -- one obtains as low or lower chance probability {\it in some bin} as that found in the lowest bin in the real dataset.  Taking this effect into account typically increases the estimation of the chance probability by a factor of a few.  For example, for the AG50 sample finding a probability of  0.02\% or lower in any 0.1\dg bin occurs in 0.09\% of the random trials, and a probability lower than 0.1\% in 0.2\dg bins occurs in 0.3\% of the cases.

It is possible that a correlation between x-ray clusters and UHECRs arises because massive galaxy clusters host objects which produce UHECRs but which are not found exclusively there.  Previous works have reported on an intriguing degree of correlation between BL Lacterae objects and UHECRs\cite{TT:BLLacs,HR:BLLacs}, so we looked to see if any of the UHECR directions correlating with an x-ray cluster is a direction which correlates with a BL Lacertae object.  We find that only one UHECR direction correlates with both a BL Lac and an x-ray cluster.  Active galactic nuclei are also considered to be possible sources of UHECR so we checked whether the UHECR correlate with x--ray selected AGN using a catalog of 99 x--ray extended sources recognized not to be clusters (mainly AGN, galaxies and stars). No significant correlation is found between the AGASA data and this sample (only one, at a separation of 1.6$^\circ$).  Therefore we conclude that the correlations between AGASA UHECR and galaxy clusters does not seem to be driven by the presence of BL Lac or AGN within the galaxy clusters.   

All three UHECR datasets show a positive correlation with x-ray clusters, so we now explore whether the observed correlations fit into a consistent picture in which the angular separation between the x-ray cluster source and the observed UHECR direction is due to a combination of resolution and magnetic smearing and deflection.  The AGASA resolution varies considerably from event-to-event depending on whether the event is well-contained or simply inside the detector array.  The calculated AGASA average angular resolution decreases rapidly with increasing energy, as is shown in Fig. 1 of \cite{AGASAclusters}:  the radius of the cone containing 68\% of the probability, $\sigma_{68}^{\rm res}$ is about 2\dg at 40 EeV and about 1.3\dg at 50 EeV, although there is some evidence that the actual average resolution is better than the calculated value (M. Teshima, private communication). The corresponding HiRes error radius, $\sigma_{68}^{\rm res} \approx 0.6^\circ$, varies little from event to event or with energy.  Making the hypothesis that the observed correlations are real, one can interpret the data as follows:\\
$\bullet$  The onset of the plateau in the radius dependence of excess correlations in the HR dataset can be used to get a crude estimate the magnetic smearing radius as follows.  Take 2\dg to be the radius containing 95\% of the genuine correlations in the HR dataset, and take the HiRes resolution smearing and the magnetic smearing to be gaussian.  In that case, the magnetic and resolution smearing add in quadrature and (in obvious notation) $\sigma_{95}^{\rm res} = 1.62 \sigma_{68}^{\rm res}$, or 1\dg for HiRes.  Thus the average value over this dataset of $\sigma_{95}^{\rm mag}$ is about 1.7\dg.  \\ 
$\bullet$ The median values for the HR, AG40 and AG50 datasets of $E^{-1}$ are $\approx 0.05,~0.019,~0.014~{\rm EeV}^{-1}$ respectively, so the "predicted" magnetic smearing radii for AG40 and AG50 are about 0.6\dg  and 0.5\dg which is small compared to the resolution-smearing for AGASA and roughly consistent with the observed onset of the plateaux in excess events for those datasets.\\   

The number of excess correlations, above the number expected by random chance, divided by the total number of UHECR events in the dataset, is an estimate of the fraction of UHECR events produced by x--ray flux-selected galaxy clusters.  This is about $\frac{4}{21},~\frac{6}{36},~\frac{12}{143}$ for the 3 datasets, within statistics allowing for this interpretation to be consistent, with a common fraction of about 12\%.  Multiplying by a factor of 2 because NORAS is only about 50\% complete, we see that galaxy clusters with x-ray emission satisfying the NORAS selection criteria, could account for of order 25\% of the UHECRs.  The analogous search for BL Lacs\cite{TT:BLLacs, HR:BLLacs} found an excess of about 9-above-expected-random correlations, using all 271 HiRes UHECRs (possible since the BL Lac catalog is an all-sky catalog), or about 3\% of the sample. 

If both the UHECR-BL Lac correlation and the correlation we report here between UHECR and massive galaxy clusters not containing BL Lacs are valid, it could be evidence of multiple acceleration mechanisms for UHE cosmic rays.  The very small angle correlation between HiRes UHECRs above 10 EeV and BLLacs, which was interpreted as evidence for a neutral component to the UHECRs, is not so evident when correlating x-ray clusters with the HiRes dataset.  This may be that statistics are still inadequate to make such comparisons, but it could also follow on physical grounds:  The BL Lac environment has a higher density of particles and radiation than x-ray clusters, so that a larger fraction of primary charged UHECRs would interact -- in the BL Lac but not massive cluster environment -- to produce $\pi^0$'s which in turn produce UHE photons.   

This evidence for a correlation between UHECRs and massive galaxy clusters can be tested with independent datasets in the near future.  HiRes anticipates accumulating 70\% more data by the end of the experiment in March 2006. The NORAS catalog is in the process of being updated,  and an x--ray cluster  catalog (CIZA) at $|b|<20^\circ$ is currently in preparation; when these are available the remaining events in the published AGASA and HiRes datasets can be studied.   Without the angular cuts imposed by the NORAS spatial restriction, the UHECR datasets used here would have been about 30\% larger.  Auger will provide an opportunity to test the correlations with  Rosat Southern Hemisphere x-ray catalogs.

E.P. is an ADVANCE fellow (NSF grant AST-0340648), also supported by NASA grant NAG5-11489; GRF's research on this project was supported by NSF-PHY-0401232.  We wish to thank Hans Bohringer and David Helfand for helpful conversations and acknowledge use of the Healpix package.
\bibliography{UH,CR}

\end{document}